\documentclass[aps,pra,floatfix,twocolumn,superscriptaddress,showpacs,10pt]{revtex4-1}
\usepackage{amssymb, amsmath,color,mciteplus,graphicx,subfigure}
\usepackage{calc,epsfig,epstopdf,color,mciteplus,bm,mathrsfs}
\usepackage{times}
\usepackage{float}
\usepackage{lipsum}

\begin{document}

\title{Ultrafast Holonomic Quantum Gates}

\author{Pu Shen}
\affiliation{Guangdong Provincial Key Laboratory of Quantum Engineering and Quantum Materials, 
and School of Physics\\ and Telecommunication Engineering, South China Normal University, Guangzhou 510006, China}

\author{Tao Chen}
\affiliation{Guangdong Provincial Key Laboratory of Quantum Engineering and Quantum Materials, 
and School of Physics\\ and Telecommunication Engineering, South China Normal University, Guangzhou 510006, China}

\author{Zheng-Yuan Xue}\email{zyxue83@163.com}
\affiliation{Guangdong Provincial Key Laboratory of Quantum Engineering and Quantum Materials, 
and School of Physics\\ and Telecommunication Engineering, South China Normal University, Guangzhou 510006, China}
\affiliation{Guangdong-Hong Kong Joint Laboratory of Quantum Matter, and Frontier Research Institute for Physics,\\ South China Normal University, Guangzhou 510006, China}

\date{\today}

\begin{abstract}
Quantum computation based on geometric phase is generally believed to be more  robust against certain errors or noises than the conventional dynamical strategy. However, the gate error caused by the decoherence effect is inevitable, and thus faster gate operations are highly desired.  Here, we propose a nonadiabatic holonomic quantum computation (NHQC) scheme with detuned interactions on $\Delta$-type three-level system, which combines the time-optimal control technique with the time-independent detuning adjustment to further accelerate universal gate operations, {so that the gate-time can be greatly shortened within the hardware limitation}, and thus high-fidelity gates can be obtained. Meanwhile, our numerical simulations show that the gate robustness is also stronger than previous  schemes. Finally, we  present an implementation of our proposal on  superconducting quantum circuits, with a decoherence-free subspace encoding, based on the experimentally demonstrated parametrically tunable coupling technique, which  simplifies previous investigations. Therefore, our protocol provides a more promising alternative for future fault-tolerant quantum computation.
\end{abstract}

\maketitle

\section{Introduction}

Quantum computers are potentially capable of solving  certain problems that are   hard for classical computers \cite{1}, and  thus the research of implementing a quantum computer has being attracted much attention for several decades. Nowadays, elementary  quantum manipulation has been verified on various systems \cite{2,3,4,5,6}, among which the superconducting quantum circuits system is one of the most promising candidates \cite{6,7,8,9,10}. However, up to now, infidelities in quantum gates induced by the decoherence effect and operational errors are still the main obstacles for physical implementation of large-scale  quantum computers. Therefore, due to their inherent noise-resilient  characteristic,  geometric phases \cite{11,12,13} are naturally used to induce robust quantum gates, and this  strategy  provides a  promising alternative for large-scale fault-tolerant quantum computation. Specifically, there are many explorations of quantum computation  using the adiabatic Abelian  \cite{14} and non-Abelian geometric phases \cite{15}. And, the adiabatic Abelian case has also been generalized to the  nonadiabatic Abelian cases \cite{16,17,18,19,20}. Recently, the research in this field is extended to investigate the nonadiabatic holonomic quantum computation (NHQC) \cite{21,22}, based on the non-Abelian geometric phase, as it is faster than the adiabatic case and can naturally be used to construct universal quantum gates. At present, the better robust performance of geometric quantum computation has been theoretically investigated \cite{20, liu2019, lisai2020, xujing2020, zhoujian2021, lisai2021} and some of them have already been experimentally demonstrated \cite{yan2019, zhu2019, xu2020, ai2020, ai2021}.

Besides, the other key indicator for a better quantum gate is the gate-fidelity. However, due to the   cyclical evolution requirement, the needed time for previous nonadiabatic  geometric quantum gates are longer than conventional dynamical schemes. Therefore, one needs to reduce the gate operation error caused by the decoherence effect by decreasing the needed gate-time. Thus,  quantum gates with the time-optimal control (TOC)  \cite{23,24,25,26} is a natural solution, where a unitary gate can be obtained  with the shortest time, by solving the quantum brachistochrone equation. In the quest of faster holonomic quantum gates, an unconventional NHQC scheme combining with TOC technique has been proposed \cite{BNHQC} based on the resonant $\Lambda$-type three-level systems and be extended \cite{28} to the case of using detuned interaction and encoded logical qubits. In these two proposals, they can achieve shorter gate-time and stronger robustness, as experimentally demonstrated in Ref. \cite{tocexp}, compared to the conventional NHQC scheme in the single-loop way \cite{34,35,36}. Then, it is naturally to ask: "how fast can holonomic quantum gates eventually be?

Here,  we propose a  NHQC scheme based on detuned $\Delta$-type three-level system, which   generalizes the previous schemes \cite{BNHQC, 28}. Different from  previous schemes, our scheme can combine TOC technique with time-independent detuning adjustment, and thus can achieve further acceleration for all  holonomic rotation gates. Remarkably, {in our proposal, the gate-time can be greatly shortened within the hardware limitation} and thus it can achieve  higher gate fidelities and stronger robustness than previous schemes. {Note that, in Ref. \cite{lisai2021}, the gate-robustness  enhancement is achieved   by incorporating  the dynamical correction technique, but the disadvantage there is that the dynamical correction process will lengthen the gate-time. That is, Ref. \cite{lisai2021} obtains the enhancement  at the cost of  decreasing  the gate-fidelity, so that it is only applicable to low-decoherence quantum systems.} Moreover, we present a physical realization of our protocol, with decoherence-free subspace (DFS) encoding \cite{29,30,31}, on a two-dimensional (2D) superconducting quantum circuit. In our implementation, for the two-logical qubit gates, we only needs coupling two physical qubits, each from a logical qubit, and thus greatly simplified previous investigations \cite{22, zhu2019, xue2015, xue2016, 32}. Therefore, our scheme provides a ultrafast and implementable alternation for NHQC, and thus is promising  for future  fault-tolerant quantum computation.

\section{Accelerated holonomic quantum gates}

\subsection{General method for constructing holonomic gates}

For a general $m$-dimensional quantum system and its quantum dynamics is governed by a time-dependent Hamiltonian $\mathcal{H}(t)$, there has an arbitrary set of orthogonal basis $\{|\Psi_m(t)\rangle\}$ and the vectors of which satisfy the time-dependent Schr\"{o}dinger equation. Substituting $|\Psi_m(t)\rangle=U(t)|\Psi_m(0)\rangle$ to the time-dependent Schr\"{o}dinger equation, we can solve  the corresponding time-evolution operator as $$U(t)=\textbf{T}e^{-\mathrm{i}\int_0^t{\mathcal{H}(t')  dt'}}=\sum_m|\Psi_m(t)\rangle\langle\Psi_m(0)|,$$
where $\textbf{T}$ is the time-ordering operator. To induce a non-Abelian geometric phase, we introduce a different set of auxiliary basis $\{|\psi_k(t)\rangle\}$, which satisfies the following boundary conditions, at the initial $t=0$ and final  $t\!=\!\tau$ moments, $|\psi_k(\tau)\rangle\!=\!|\psi_k(0)\rangle\!=\!|\Psi_k(0)\rangle$. Using $|\psi_k(t)\rangle$, $|\Psi_m(t)\rangle$ can be expanded as $|\Psi_m(t)\rangle=\sum_m C_{mk}(t)|\psi_k(t)\rangle$. Substituting the expansion and the boundary conditions into the time-dependent Schr\"{o}dinger equation, the time-evolution operator can be obtained, after a time interval $[0, \tau]$, as
\begin{eqnarray} 	\label{EqUtau} U(\tau)=\sum_{ml}[\textbf{T}e^{\mathrm{i}(\textbf{A}+\textbf{K})}]_{ml}|\Psi_m(0)\rangle\langle\Psi_l(0)|,
\end{eqnarray}
where the matrix elements $K_{ml}\!=\!-\int_0^\tau\langle\psi_m(t)|\mathcal{H}(t)|\psi_l(t)\rangle dt$ and $A_{ml}=\int_0^\tau \mathrm{i}\langle\psi_m(t)\lvert\frac{d}{dt}|\psi_l(t)\rangle dt$ represent the dynamical and geometric parts of the total phase, respectively. Different from the conventional geometric restriction that the dynamical phase is strictly required to be zero, we here just require the dynamical part to hold the geometric properties \cite{BNHQC, 28}, in the form of $\textbf{K}=r\textbf{A}+\textbf{G}$ with $r$ being a proportional constant and the matrix $\textbf{G}$ depending only on the global geometric feature of the evolution path. In this way, a unconventional holonomy from Eq. (\ref{EqUtau}) can be obtained, as proofed in Ref. \cite{28}.

\begin{figure}[tbp]
	
	\includegraphics[width=0.9\linewidth]{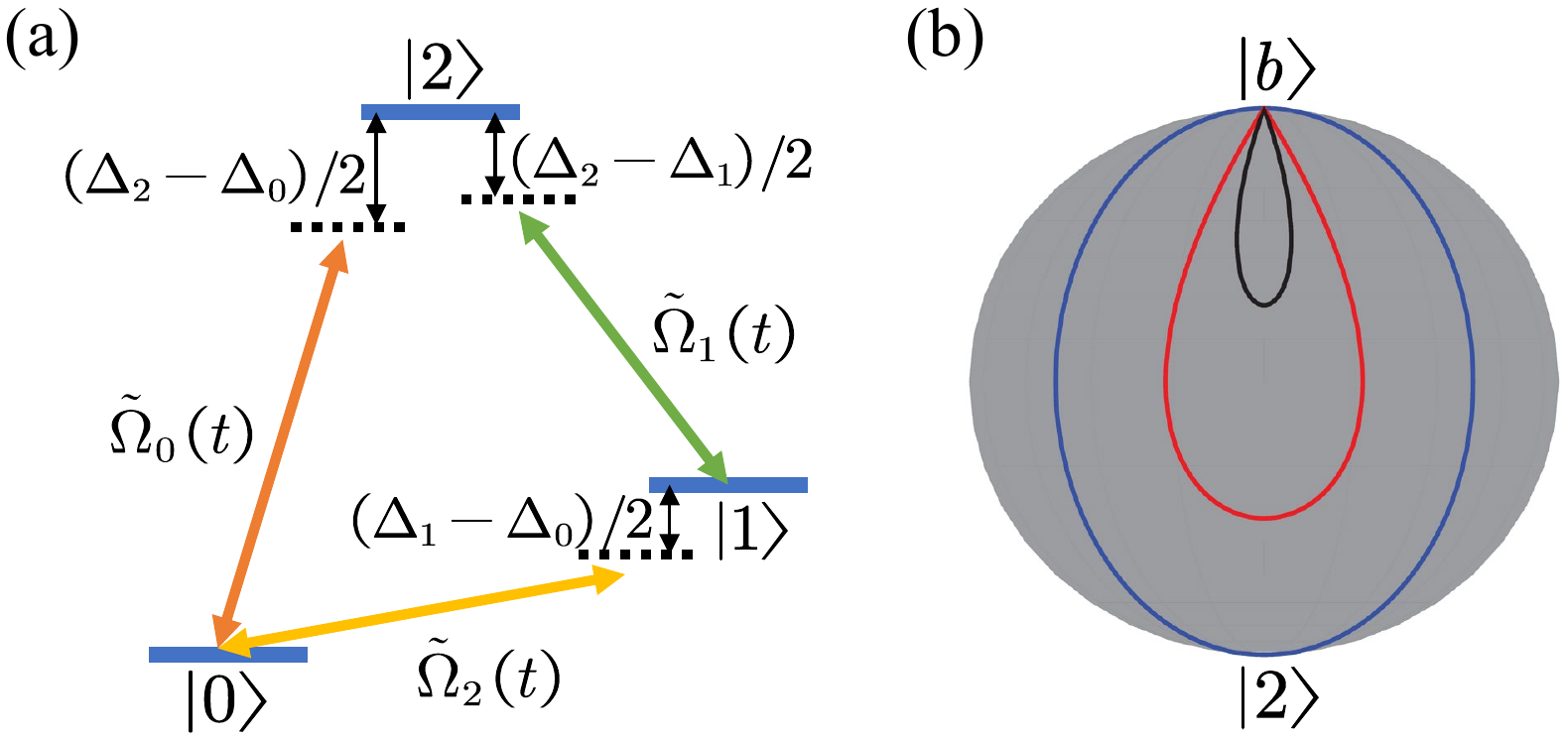}
	\caption{Illustration of our ultrafast holonomic quantum gate scheme. (a) A cyclical and detuned coupled  $\Delta$-type three-level system, where two ground states serve as our qubit-states. (b) Illustration of the evolution paths of the $\pi/2$   rotating operation  around an arbitrary axis, on the X-Z plane projection of the Bloch sphere, for  the conventional single-loop NHQC scheme (blue line) and our proposal with $\Delta_2/\Omega = 0$ (red line) and $|\Delta_2/\Omega|= 1/2$ (black line).}
	\label{Figure.1}
\end{figure}

\subsection{Application to $\Delta$-type three-level system with TOC}
For the pursuit of high-fidelity gate operation, the gate-time is required to be as short as possible, to reduce the infidelity caused by the decoherence and error effects. Thus, considering a $\Delta$-type three-level system, we next utilize the TOC technique \cite{23,24,25,26} to find the shortest evolution path to accelerate the quantum gates.

As shown in Fig. \ref{Figure.1}(a), a $\Delta$-type three-level system has two ground states $|0\rangle$ and $|1\rangle$ which are used as our qubit-states, and an excited state $|2\rangle$ is served as an auxiliary state. There exists three sets of transitions, i.e., $|0\rangle\!\leftrightarrow\!|2\rangle$, $|1\rangle\!\leftrightarrow\!|2\rangle$ and $|0\rangle\!\leftrightarrow\!|1\rangle$, driven by three microwave fields  $\tilde{\Omega}_i=\Omega_i(t)\cos[\upsilon_it-\phi_i(t)]$ with $i\in\{0, 1, 2\}$, where the driving frequency, amplitude and initial phase are denoted by $\upsilon_i$, $\Omega_i(t)$, and $-\phi_i(t)$, respectively. Assuming $\hbar\!=\!1$ hereafter, the Hamiltonian of the driven quantum system be written as
\begin{eqnarray}
	\label{initialHt}
	\mathcal{H}(t)&&=\!\sum_{n=0,1,2}\!\!{\omega_n|n\rangle\langle n|}\notag\\
	&&\!+\!\left\{\tilde{\Omega}_0(t)|2\rangle\langle 0|
	\!+\!\tilde{\Omega}_1(t)|2\rangle\langle 1|\!+\!\tilde{\Omega}_2(t)|1\rangle\langle 0|\!+\!\mathrm{H.c.}\right\}, \quad
\end{eqnarray}
where  $\omega_n$ represents the level energy. Then, we apply a unitary transformation $U_0(t)=\exp{{(-\mathrm{i}\sum_{n=0}^2\omega'_n|n\rangle\langle n|t)}}$, with $\omega'_2-\omega'_0=\upsilon_0$, $\omega'_2-\omega'_1=\upsilon_1$, and $\omega'_1-\omega'_0=\upsilon_2$,  on the above Hamiltonian, {under the rotating wave approximation}, the transformed Hamiltonian reads
\begin{eqnarray}
	\label{Htrans}
	 \mathcal{H}_{\textrm{t}}(t)&&={\frac{1}{2} \sum_{n=0}^{2} \Delta_n|n\rangle\langle n|}
	 \!+\!\frac{1}{2}\left\{\Omega_0(t)e^{\mathrm{i}\phi_0(t)}|2\rangle\langle 0|\right. \notag\\
     &&\!+\!\left.\Omega_1(t)e^{\mathrm{i}\phi_1(t)}|2\rangle\langle 1|\!+\!\Omega_2(t)e^{\mathrm{i}\phi_2(t)}|1\rangle\langle 0|\!+\!\mathrm{H.c.}\right\},
\end{eqnarray}
where $\Delta_n\!=\!2(\omega_n\!-\!\omega'_n)$. {By setting $\Omega(t)=\sqrt{\Omega^2_0(t)+\Omega^2_1(t)}$, $\theta\!=\!2\tan^{-1}[\Omega_0(t)/\Omega_1(t)]$, $\Delta_0\!=\!-\Delta_2 \sin^2(\theta/2)$, $\Delta_1\!=\!-\Delta_2 \cos^2(\theta/2)$ and $\Omega_2(t)\!=\!-\Delta_2 \sin(\theta/2)\cos(\theta/2)$}, in the dressed-state representation, the dynamics of the quantum system can be regarded as a detuned coupling  between the bright state {$|b\rangle\!\!=\!\sin(\theta/2)e^{-\mathrm{i}\phi}|0\rangle\!+\cos(\theta/2)|1\rangle$}, with {$\phi\equiv\phi_2(t)=\!\phi_0(t)\!-\!\phi_1(t)$} being a constant, and the auxiliary state $|2\rangle$, i.e.,	$\mathcal{H}_{\textrm{eff}}(t)=\mathcal{H}_0+\mathcal{H}_c(t)$ with
\begin{subequations}  	\label{Heff}
\begin{eqnarray}
\mathcal{H}_0&=& -\frac{\Delta_2}{2}(|b\rangle\langle b|\!-\!|2\rangle\langle 2|),
\end{eqnarray}
\begin{eqnarray}
\mathcal{H}_c(t)&=& \frac{\Omega(t)}{2}(e^{-\mathrm{i}\phi_1(t)}|b\rangle\langle 2|\!+\!\mathrm{H.c.}).
\end{eqnarray}
\end{subequations}
In addition, there has a dark state $|d\rangle\!=\!\cos(\theta/2)|0\rangle\!-\!\sin(\theta/2)e^{\mathrm{i}\phi}|1\rangle$, which is decoupled from the dynamics.

\begin{figure*}[tbp]
	
	\includegraphics[width=0.75\linewidth]{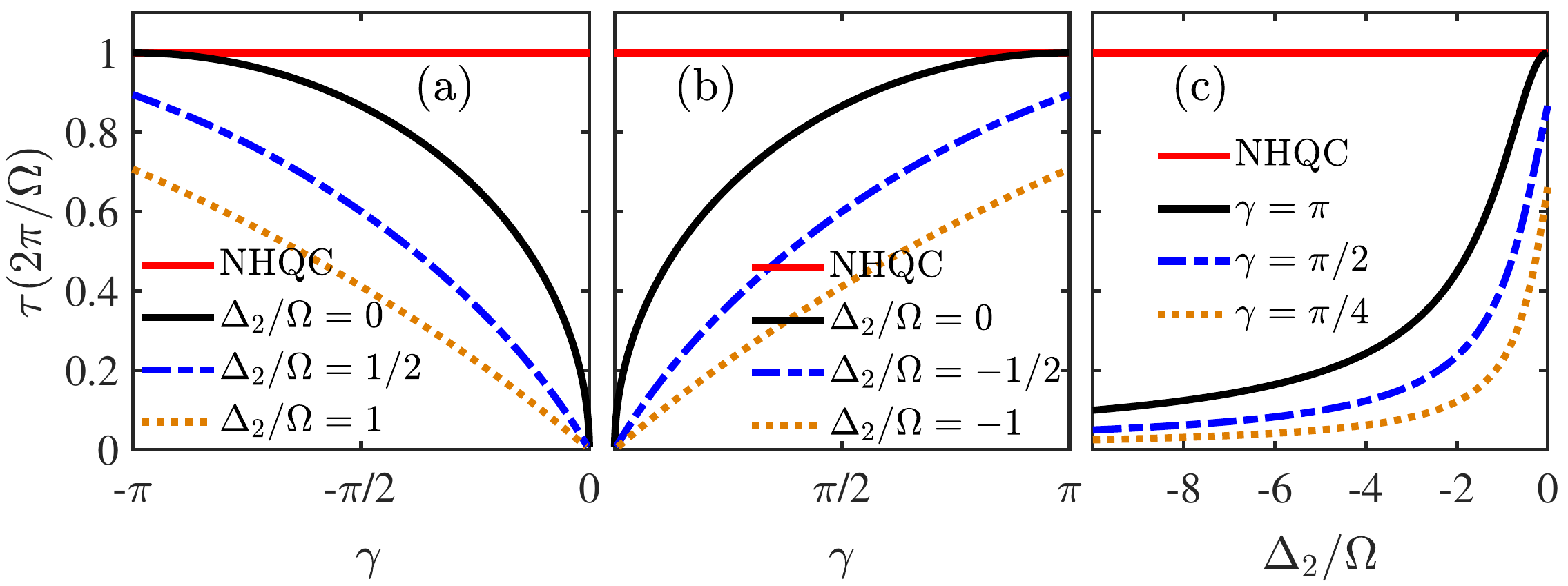}
	\caption{Illustration of the needed gate-time for our scheme ($\Delta_2/\Omega>0$), previous NHQC  schemes with TOC ($\Delta_2/\Omega=0$), and the conventional single-loop NHQC scheme. When $\Delta_2/\Omega$ is positive (a) or negative (b), the  gate-time $\tau$ is plotted  with respect to the rotation angle $\gamma$, which indicates that increasing the ratio  $|\Delta_2|/\Omega$ can speed up the gates. (c) For different rotation angles $\gamma$, {all the gate-time can be greatly  shortened, when increasing $|\Delta_2|/\Omega$.}  }
	\label{Figure.2}
\end{figure*}

Note that the positive and negative  $\Delta_2$ correspond to the negative and positive detuning of two transitions between $|0\rangle\!\leftrightarrow\!|2\rangle$ and $|1\rangle\!\leftrightarrow\!|2\rangle$. In the following, we will show that this additional parameter enable us to accelerate the induced holonomic quantum gates, { and the gate-time can be greatly shortened within the hardware limitation.} This is the main difference between this work and previous ones \cite{BNHQC, tocexp, 28,  34,35,36}, and can be explained intuitively as following. Any target gate is act on the qubit-states, and thus the main contribution of the gates will be attributed to the  $|0\rangle \leftrightarrow |1\rangle$ transition, while other transitions are introduced to make the gate to be holonomic.

We now consider introducing the TOC into the construction of arbitrary holonomic quantum gates. In this case, the engineering of the Hamiltonian in Eq. (\ref{Heff}) has two constraints \cite{26,BNHQC,28}:   (i)  the strength  of the driving fields cannot be infinite, due to the finite energy bandwidth of a quantum system; and (ii) the Hamiltonian of a realistic quantum device usually takes a given form, so that only limited quantum trajectory can be realized. Mathematically, for our system, these two constraints correspond to
\begin{flalign}
	&(\mathrm{i})\quad l_1(\mathcal{H}_c)=\frac{1}{2}[\textrm{Tr}(\mathcal{H}_c^2(t))-\frac{1}{2}\Omega^2(t)]=0,& \nonumber \\
	&(\mathrm{ii})\quad l_2(\mathcal{H}_c)=\textrm{Tr}[\mathcal{H}_c(t)\tilde{\sigma}_z]=0,&
\end{flalign}
where $\tilde{\sigma}_{x,y,z}$ is Pauli operators in the subspace of $\{|b\rangle,|2\rangle\}$. Furthermore, the parameters of the microwave fields can be determined by solving the quantum brachistochrone equation of $\partial F/\partial t\!=\!-\mathrm{i}[\mathcal{H}_{\textrm{eff}}(t),F]$, where $F\!=\!\partial (\sum_{j=1,2}\lambda_j l_j(\mathcal{H}_c))/\partial \mathcal{H}_c$ with $\lambda_j$ being the Lagrange multiplier. In order to obtain the fastest gate, the parameters of the driving fields are required to meet the condition of
\begin{eqnarray}
	\phi_1(t)\!=\!\int_0^t{[\Delta_2\!+\!c\Omega(t')]dt'},
\end{eqnarray}
after defining $\lambda_1\!=\!1/\Omega(t)$ and $\lambda_2\!=\!c/2$ being a constant. {Here, we take $\Omega(t)\!=\!\Omega$, i.e., the square pulse case, where $\Omega(t)$ will be at its maximum value within the hardware limitation all the time, so that it can reach a certain pulse area in the shortest time.} In this case, the constraint on $\phi_1(t)$ under TOC will  reduce to $\phi_1\!(t) =\eta t$ with $\eta= (\Delta_2\!+\!c\Omega)$ being a constant. Therefore,   at the final time $\tau$, the resulting evolution operator on the basis of $\{|b\rangle, |2\rangle\}$ can be obtained as
\begin{eqnarray}
	\label{Ube}
	U(\tau)\!=\!e^{-\mathrm{i}\frac{1}{2}\eta \tau\tilde{\sigma}_z}
	\!\left(
	\begin{array}{cccc}
		\!\!\cos\xi\!+\!\mathrm{i}\sin\xi \cos\frac{\chi}{2}            & \!-\mathrm{i}\sin\xi \sin\frac{\chi}{2} \\
		\!-\mathrm{i}\sin\xi \sin\frac{\chi}{2} & \!\cos\xi\!-\!\mathrm{i}\sin\xi \cos\frac{\chi}{2}\!\!
	\end{array}
	\right), \notag\\
\end{eqnarray}
where $\xi\!=\!\!\sqrt{\Omega^2\!+\!(\eta\!+\!\Delta_2)^2}\tau/2$ and $\chi\!=\!2\tan^{-1}[\Omega/(\eta\!+\!\Delta_2)]$. Furthermore, to ensure  the cyclic evolution condition is satisfied, we here set $\xi\!=\!\pi$. {In this way,  in the subspace of $\{|b\rangle,|d\rangle,|2\rangle\}$, the targeted evolution operator will be
\begin{eqnarray}
	\label{Ubd}
	U(\tau)=e^{-\mathrm{i}\gamma}|b\rangle\langle b|+|d\rangle\langle d|+e^{\mathrm{i}\gamma}|2\rangle\langle 2|,
\end{eqnarray}
where the total phase $\gamma\!=\!\pi\!+\!\frac{1}{2}\eta\tau$. In the computational subspace $\{|0\rangle,|1\rangle\}$, the above evolution operator reads}
\begin{eqnarray}
\label{U01}
U(\tau)=e^{-\mathrm{i}\frac{\gamma}{2}}e^{-\mathrm{i}\frac{\gamma}{2}\vec{n}\cdot\vec{\sigma}},
\end{eqnarray}
which is a rotating operation around the $\vec{n}\cdot\vec{\sigma}$ axis by $\gamma$ angle, where $\vec{n}=(\sin\theta \cos\phi, \sin\theta \sin\phi, -\cos\theta)$ and $\vec{\sigma}=(\sigma_x, \sigma_y, \sigma_z)$ with $\sigma_{x,y,z}$ being Pauli operators in the  qubit subspace of $\{|0\rangle,|1\rangle\}$. In addition, the total phase $\gamma\!=\!\pi\!+\!\frac{1}{2}\eta\tau\!=\!\boldsymbol{\textbf{G}_{11}}$ can be regarded as an unconventional geometric phase that satisfies the definition in Eq. (\ref{EqUtau}). Therefore, when setting the parameters $(\gamma,\theta,\phi)$ to $(\theta_x,\pi/2,\pi)$, $(\theta_y,\pi/2,\pi/2)$ and $(\theta_z,\pi,\pi)$, arbitrary holonomic X-, Y-, Z-axis rotating gates $R_x(\theta_x)$, $R_y(\theta_y)$ and $R_z(\theta_z)$ can be realized, respectively. In the Bloch sphere, as shown in Fig. \ref{Figure.1}(b), the evolution path of our scheme is shorter than that of single-loop schemes \cite{34,35,36}, and the evolution path can be further shortened by increasing the detuning $|\Delta_2|$, which can also be shorter that  previous TOC schemes \cite{BNHQC, tocexp, 28}.

\subsection{Acceleration effect}

For the gate-time in conventional NHQC schemes, it is limited by the fact that all the gates need strictly a same time of $\tau_c\!=\!2\pi/\Omega$, even for a small rotation-angle operation \cite{33}. Here, by uniting the equations $\gamma\!=\!\pi\!+\!\frac{1}{2}\eta\tau$ and $\xi\!=\!\pi$, we find that the relationship between the gate-time $\tau$ and the rotation angles $\gamma$ around different axes can all be solved as
\begin{eqnarray}
	\label{time}
	\tau=\tau_c\sqrt{1-(\gamma/\pi-1)^2(1+\Delta_2/\eta)^2},
\end{eqnarray}
which is obviously less than $\tau_c$. Meanwhile, in our scheme, we can further speed up the gate  by tuning the detuning $\Delta_2$ to achieve  fully accelerated arbitrary holonomic quantum  operations. Note that, as seeing from Eq. (\ref{Heff}), when $\Delta_2=0$, our scheme  reduces  to the resonant $\Lambda$-type three-level case, i.e., the previous NHQC schemes combining with TOC \cite{BNHQC,28}. As shown in Fig. \ref{Figure.2}(a) and \ref{Figure.2}(b), our gate-time is also considerably smaller than these schemes, even with relatively small detuning, i.e., $\Delta_2\sim \Omega$.

\begin{figure}[tbp]	
	\includegraphics[width=0.95\linewidth]{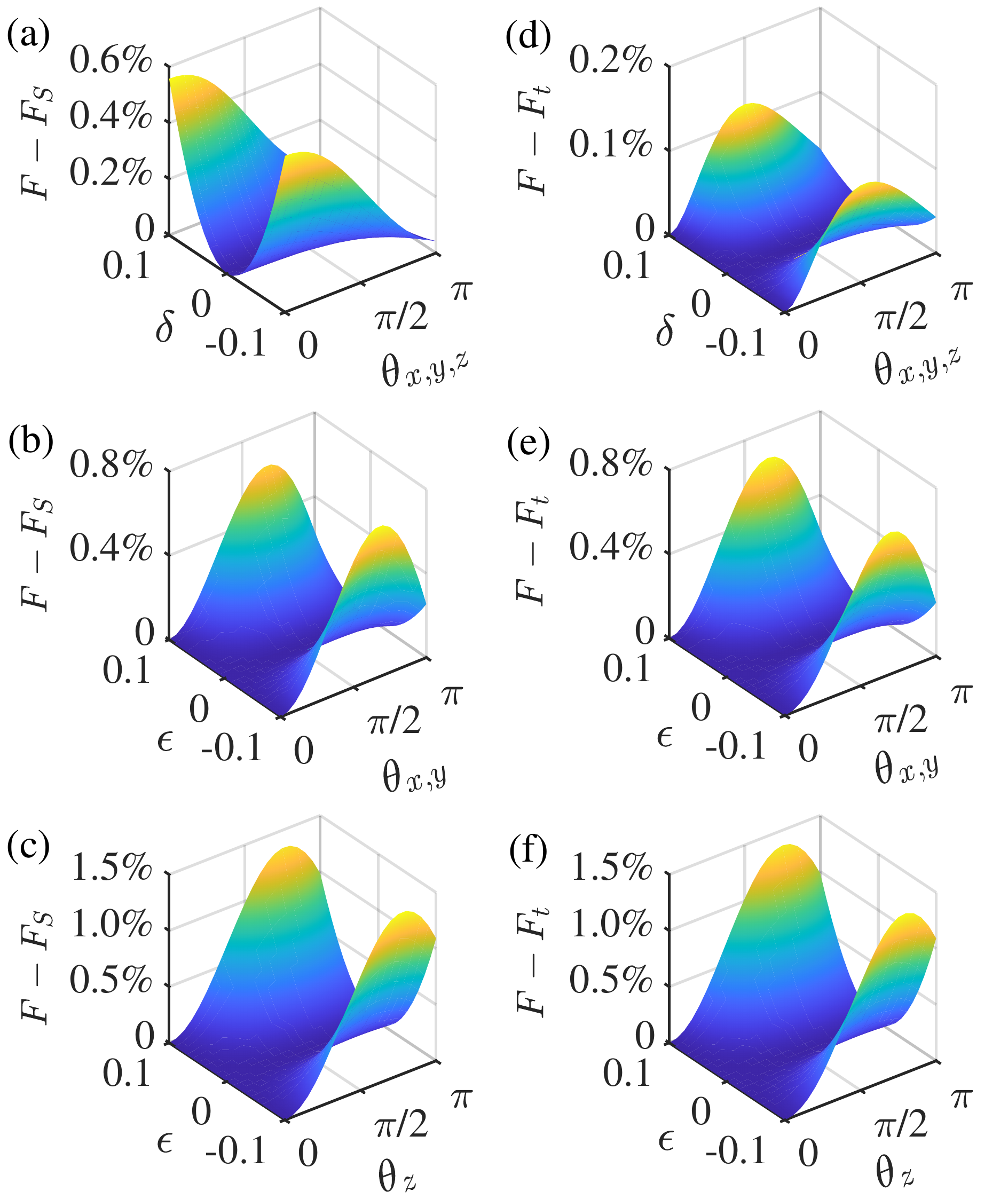}
	\caption{Comparison of the gate-robustness for different schemes. The fidelity difference between our scheme and (a) single-loop NHQC scheme  and (d) previous TOC schemes for the qubit frequency drift $\delta\times\Omega$, where  results of the three rotation gates $R_{x,y,z}(\theta_{x,y,z})$ are the same. As to the coupling strength error $\epsilon\times\Omega$, the fidelity difference for $R_x(\theta_x)$ and $R_y(\theta_y)$ are the same, as shown in (b) for  single-loop NHQC scheme  and (e) previous TOC schemes; but the $R_z(\theta_z)$ case is different, as shown in (c) for single-loop NHQC scheme and (f) previous TOC schemes. Anyway, our scheme has the best performance for all cases.}
	\label{Figure.3}
\end{figure}

Thus,  our scheme is different from previous NHQC schemes combining with TOC  \cite{BNHQC,28}, they do not have this additional parameter to  speed up arbitrary gate operations. Taking several typical rotation operations as examples, as shown in Fig. \ref{Figure.2}(c), we find that the gate-time has been greatly shortened with the adjustment of the detuning parameter $\Delta_2$. Note that, although parameter $\Delta_2$ cannot be increased infinitely experimentally, tuning it within the hardware limitation can still achieve a considerable acceleration for all holonomic gate operations. {Finally, although shorter evolution path may lead to faster quantum gates, the gate-time is also determined by how fast the path is traveled, which is also limited by the geometric conditions. Here,  we focus on optimizing the gate-time instead of the path, while the relationship between the gate-time and the gate-path is complicated.}

\section{Gate performance}

In the following, we proceed to analyze the advantage of  our gate strategy in terms of the  gate robustness and gate fidelity.
To illustrate the noise-resilient feature of our fully accelerated holonomic quantum gate against systematic errors, we numerically simulate the gate performance as a function of the qubit-frequency drift and the deviation of coupling strength. In this case, the Hamiltonian of the system in Eq. (\ref{Heff}) will change to
\begin{eqnarray}
	\label{Heff2}
	\mathcal{H}'_{\textrm{eff}}(t)&=&-\frac{\Delta'}{2}(|b\rangle\langle b|\!-\!|2\rangle\langle 2|)\notag\\
&&+\!\frac{\Omega'(t)}{2}(e^{-\mathrm{i}\phi_1(t)}|b\rangle\langle 2|\!+\!\mathrm{H.c.}),
\end{eqnarray}
where $\Delta'=\Delta_2+\delta\Omega$ and $\Omega'(t)=\Omega'=\Omega+\epsilon\Omega$. The gate fidelity is defined as $F=\textrm{Tr}(U^\dagger U')/\textrm{Tr}(U^\dagger U)$, where $U'$ represents the actual gate under these two errors. To show the superiority of our scheme, we plot the gate fidelity difference between our scheme and the conventional single-loop NHQC scehme ($F_s$) \cite{34,35,36} and previous NHQC with TOC ($F_t$) \cite{BNHQC, 28}, i.e., $F-F_s$ and $F-F_t$. Fig. \ref{Figure.3} is plotted when $\Delta_2/\Omega=-1/2$ for our scheme, one finds that, in terms of the two main error sources for general qubits, our fully accelerated holonomic scheme has stronger robustness than previous schemes. Therefore, our scheme alleviates the problem of being sensitive to systematic errors caused by the requirement of a fixed pulse area in  conventional NHQC schemes.

\begin{figure}[tbp]
	\includegraphics[width=0.95\linewidth]{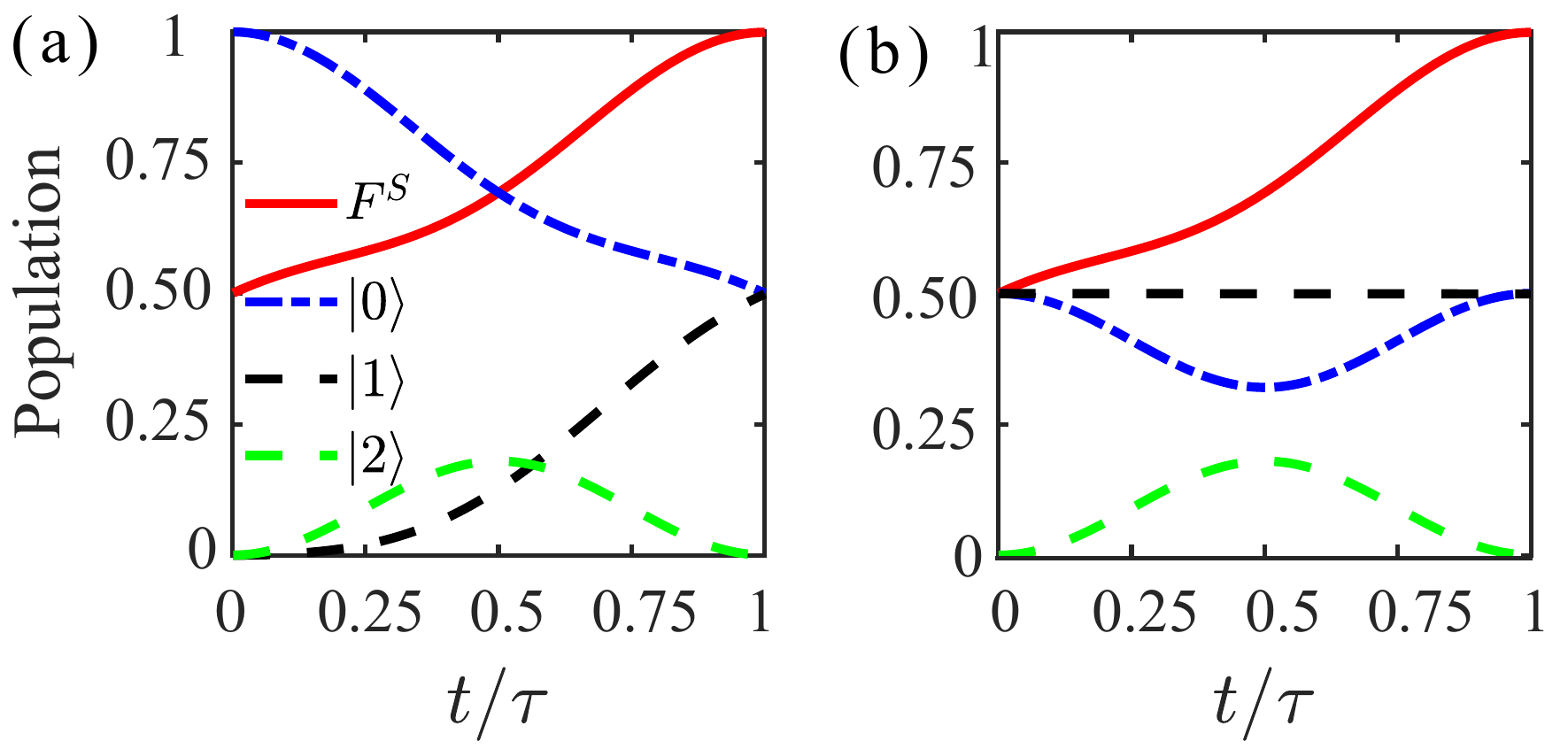}
	\caption{The logical-qubit-state population and fidelity dynamics for holonomic (a) $R_x(\pi/2)$ and (b) $R_z(\pi/2)$ gates with the initial states being $\lvert0\rangle$ and $(\lvert0\rangle+\lvert1\rangle)/\sqrt{2}$, respectively.}
	\label{Figure.4}
\end{figure}

\begin{figure}[tbp]
	\includegraphics[width=0.95\linewidth]{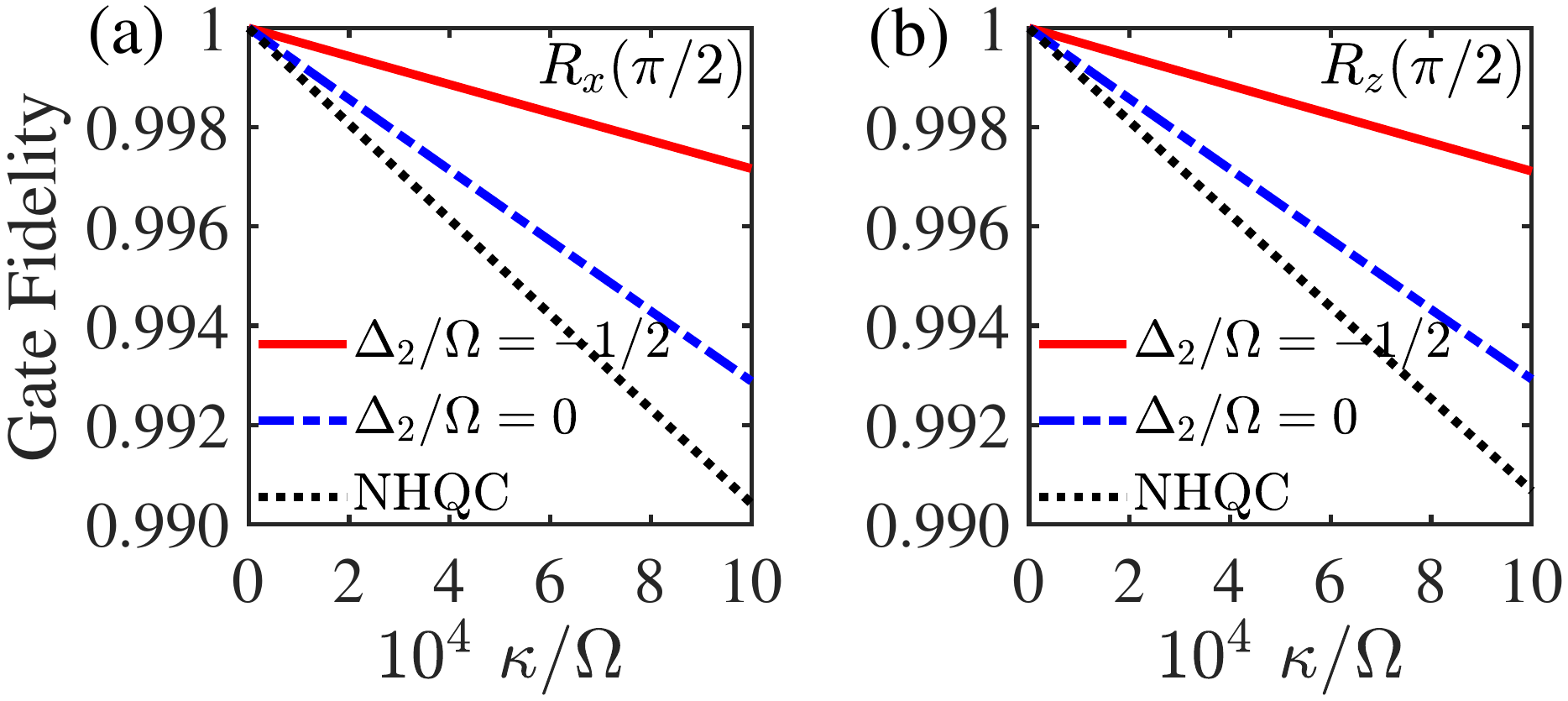}
	\caption{{Comparison of the gate fidelities for different NHQC strategies for holonomic (a)   $R_x(\pi/2)$  (b)   $R_z(\pi/2)$ gates  with different decoherence rate $\kappa$ for our scheme ($\Delta_2/\Omega\!=\!-1/2$), the previous NHQC scheme with TOC ($\Delta_2/\Omega\!=\!0$) and the conventional single-loop NHQC scheme. It is obvious that the adjustment of the detuning $|\Delta_2|$ can reduce the influence from the decoherence effect.}}
	\label{Figure.5}
\end{figure}

We next consider the influence of the decoherence effect, caused by the inevitable interaction between the quantum system and its surrounding environment,  using the Lindblad master equation of
\begin{eqnarray}
\dot{\rho}=\mathrm{i}[\rho,\mathcal{H}(t)]+\frac{\Gamma_-}{2}\mathcal{L}(S_-)
+\frac{\Gamma_z}{2}\mathcal{L}(S_z),
\end{eqnarray}
where $\mathcal{L}(A)\!=\!2A\rho A^\dagger-A^\dagger A\rho-\rho A^\dagger A$ is the Lindblad operator with $S_-=|0\rangle\langle 2| +|1\rangle\langle 2|$ and $S_z=2|2\rangle\langle 2|-|1\rangle\langle 1|-|0\rangle\langle 0|$; $\Gamma_-$ and $\Gamma_z$ are decay and dephasing rates of the qubit system respectively. We here choose  $R_x(\pi/2)$ and $R_z(\pi/2)$ gates as two typical examples for our numerical evaluation. Supposing the initial states for $R_x(\pi/2)$ and $R_z(\pi/2)$ are $|0\rangle$ and $(|0\rangle+|1\rangle)/\sqrt{2}$, respectively, where the resulting states in the ideal situation are both $|\psi_{x,z}(\tau)\rangle=(|0\rangle+\mathrm{i}|1\rangle)/\sqrt{2}$, and we evaluate these gates by the state fidelities defined by $F^S_{x,z}\!=\!\langle\psi_{x,z}(\tau)|\rho|\psi_{x,z}(\tau)\rangle$. As shown in Fig. \ref{Figure.4}, when choosing the parameters of $\Delta_2/\Omega\!=\!-1/2$ and $\Gamma_-\!=\!\Gamma_z\!=\!\kappa\!=\!4\times10^{-4}\Omega$, we can obtain   high  fidelities $99.92\%$ and $99.90\%$ for $R_x(\pi/2)$ and $R_z(\pi/2)$, respectively. {In addition, gate fidelities changing with different decoherence rate $\kappa$ are shown in the Fig. \ref{Figure.5}, where the gate fidelities are defined as $F^G_{x,z}=\frac{1}{6}\sum_{j=1}^6{_j\langle\psi'_{x,z}(\tau)|\rho|\psi'_{x,z}(\tau)\rangle_j}$ \cite{GateF}, with the six states are $|\psi'\rangle_1\!=\!|0\rangle$, $|\psi'\rangle_2\!=\!|1\rangle$, $|\psi'\rangle_3\!=\!(|0\rangle\!+\!|1\rangle)/\sqrt{2}$, $|\psi'\rangle_4\!=\!(|0\rangle\!-\!|1\rangle)/\sqrt{2}$, $|\psi'\rangle_5\!=\!(|0\rangle\!+\!\mathrm{i}|1\rangle)/\sqrt{2}$ and $|\psi'\rangle_6\!=\!(|0\rangle\!-\!\mathrm{i}|1\rangle)/\sqrt{2}$, and the ideal final states are $|\psi'_{x,z}(\tau)\rangle_j=R_{x,z}(\pi/2)|\psi'\rangle_j$.} We find that the gate-fidelity  of our scheme ($\Delta_2\!\ne\!0$) will be higher than  both previous TOC ($\Delta_2=0$) and conventional single-loop NHQC schemes. Thus, besides the gate robustness merit, our scheme also provides better suppression of the decoherence induced gate-infidelity.

\section{Physical realization on superconducting quantum circuits}

The cyclical interaction used in our protocol seems to be bizarre,  as natural atoms usually do not permit this transition configuration. However, for artificial superconducting atoms, there does exist such transition configuration \cite{cyclical1, cyclical2}, e.g., superconducting flux qubits.  But, the coherence quality is relatively low, comparing with superconducting transmon qubits \cite{transmon}. Considering all the merits and disadvantages, we here present our implementation on a 2D superconducting quantum circuit with transmon qubits. In our implementation, we also incorporate the DFS encoding technique, where effective three-level structure can be formed in the single-excitation subspace, and the cyclical coupling among three qubits can also be induced and tuned, using the experimentally demonstrated parametric coupling technique \cite{parametric1, parametric2, parametric3}.

\begin{figure}[tbp]	
	\includegraphics[width=0.8\linewidth]{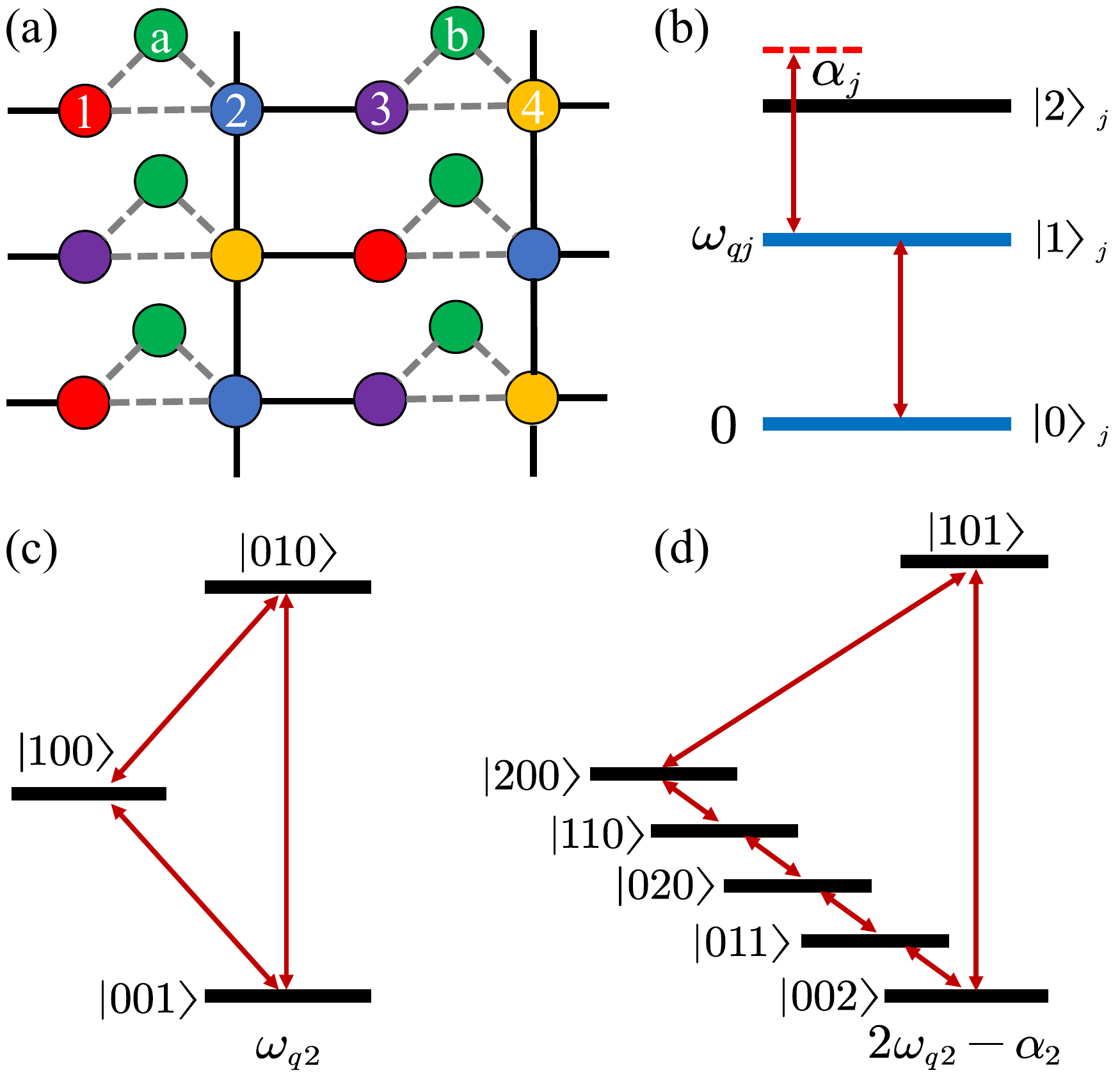}
	\caption{{(a) The proposed 2D superconducting  circuit for  implementing our scheme in the DFS,    where each logical qubit is encoded by three capacitive coupled  transmon qubits (filled circles), e.g., $\{\mathrm{T}_1,\mathrm{T}_a,\mathrm{T}_2\}$ or $\{\mathrm{T}_3,\mathrm{T}_a,\mathrm{T}_4\}$, with different colors indicate different frequencies. The grizzle dashed line represents the capacitive couplings among transmons in a logical unit and the solid line represents the capacitive coupling between two logical units. (b) Three energy levels are considered for each transmon qubit $\mathrm{T}_{j}$, with $\alpha_j$ being the anharmonicity. When implementing a single-logical-qubit gate on a logic-qubit unit $\{\mathrm{T}_1,\mathrm{T}_a,\mathrm{T}_2\}$, the quantum dynamics of the system involves two independent subspaces, i.e.,  (c) single-excitation  and (d) double-excitation subspaces.}}
	\label{Figure.6}
\end{figure}

\subsection{Single-qubit gates}

We firstly outline  our implementation of single-qubit gates. As shown in Fig. \ref{Figure.6}(a), three capacitively coupled transmon qubits are used to encode a DFS logical qubit, where the coupling strength  between transmon qubit $\mathrm{T}_i$ and $\mathrm{T}_j$ is denoted by $g_{ij}$. Only three energy levels are considered for each transmon qubit $\mathrm{T}_{j}$, where the transition frequency of the lowest two energy levels is $\omega_{qj}$ and $\alpha_j$ being the anharmonicity, as shown in Fig. \ref{Figure.6}(b). For a logical unit $\mathrm{T}_{1,a,2}$, besides an auxiliary state $|e\rangle_\mathrm{L}=|010\rangle_{1a2}$, the corresponding logical subspace is $S_1=\{ |100\rangle_{1a2} =|0\rangle_\mathrm{L}, |001\rangle_{1a2}\!=\!|1\rangle_\mathrm{L}\}$. {Assuming the initial state of the other two physical qubits are both in their ground states, an   arbitrary state on physical qubit 1, e.g., $|\psi\rangle_1=a|1\rangle_{1}+b|0\rangle_{1}$, can be encoded into the logical qubit subspace, i.e., ,
$|\psi\rangle_L=a|0\rangle_{L}+b|1\rangle_{L}$ as following.  First, a NOT gate is applied on the qubit 2, and then a CNOT gate is applied on qubits 1 and 2, with qubit 1 being the control qubit. Meanwhile, the decoding circuit can be obtained by inverse these two steps. To manipulate logical-qubit states,  the quantum system is proposed to be cyclically coupled as
\begin{eqnarray}
	\label{H1t}
	\mathcal{H}_1(t)&&=\!\sum_{j=1,a,2}{[\omega_{qj}|1\rangle_j\langle 1|+(2\omega_{qj}-\alpha_j)|2\rangle_j\langle 2|]}\notag\\
	&&+g_{1a}(\sigma_{1}+\sigma^\dagger_{1})\otimes(\sigma_{a}+\sigma^\dagger_{a})    \notag\\
	&&+g_{a2}(\sigma_{a}+\sigma^\dagger_{a})\otimes(\sigma_{2}+\sigma^\dagger_{2}) \notag\\
	&&+g_{12}(\sigma_{1}+\sigma^\dagger_{1})\otimes(\sigma_{2}+\sigma^\dagger_{2}),
\end{eqnarray}
where $\sigma_{j}\!=\!A_{j}\!+\!B_{j}$ is lower operator for transmon $\mathrm{T}_j$, with $A_{j}\!=\!|0\rangle_{j}\langle 1|$ and $B_{j}\!=\!\sqrt{2}|1\rangle_{j}\langle 2|$. Not that, to show the influence of the second excited state of  transmon qubits, we  also include the $|2\rangle$ state in the above Hamiltonian.}

To make the coupling among transmon qubits  tunable, we apply drive fields on the transmon qubits $\mathrm{T}_1$ and $\mathrm{T}_2$, i.e.,  $\omega_{qj}(t)=\omega_{qj}+\varepsilon_{j}\sin(\nu_{j}t +\varphi_{j}(t))$, where $\varepsilon_{j}$, $\nu_{j}$ and $\varphi_{j}(t)$ is the amplitude, frequency and time-dependent phase, respectively. Then, to conceal the time-dependence of the driving fields, a unitary transformation
\begin{eqnarray}
	\label{U1t}
	U_1(t)&&=e^{\!-\!\mathrm{i}\sum_{j=1,a,2}{[(\omega_{qj}\!-\!\Delta'_j)|1\rangle_j\langle 1|\!+\!(2\omega_{qj}\!-\!\alpha_j\!-\!\Delta'_j)|2\rangle_j\langle 2|]}t}\notag\\ &&\times e^{\mathrm{i}\sum_{j=1,2}{\beta_j\cos(\nu_{j}t\!+\!\varphi_j(t))(|1\rangle_j\langle 1|\!+\!2|2\rangle_j\langle 2|)}},
	\end{eqnarray}
is applied to Eq. (\ref{H1t}), where $\beta_j=\varepsilon_{j}/\nu_{j}$ and $\Delta'_j$ will be set later. Under the rotating wave approximation, {in the subspace of $\{|0\rangle, |1\rangle, |2\rangle\}$ of all the transmon qubits,} the transformed Hamiltonian reads
\begin{eqnarray}
	\label{HT1t}
	\mathcal{H}^T_1(t)&&=\!\sum_{j=1,a,2}{\Delta'_j(|1\rangle_j\langle 1|\!+\!|2\rangle_j\langle 2|)}\notag\\
	&&+\{g_{1a}[A^\dagger_{1}A_{a}e^{\mathrm{i}\Delta'_{1a}t} \!+\!A^\dagger_{1}B_{a}e^{\mathrm{i}(\Delta'_{1a}\!+\!\alpha_a)t}\notag\\
	&&+B^\dagger_{1}A_{a}e^{\mathrm{i}(\Delta'_{1a} -\!\alpha_1)t}]\!\times\!e^{-\mathrm{i}\beta_1\cos(\nu_1t\!+\!\varphi_1(t))}\notag\\
	&&+g_{a2}[A^\dagger_{a}A_{2}e^{\mathrm{i}\Delta'_{a2}t} +\!A^\dagger_{a}B_{2}e^{\mathrm{i}(\Delta'_{a2} +\!\alpha_2)t}\notag\\
	&&+B^\dagger_{a}A_{2}e^{\mathrm{i}(\Delta'_{a2} -\!\alpha_a)t}]\!\times\!e^{\mathrm{i}\beta_2\cos(\nu_2t\!+\!\varphi_2(t))}\notag\\
	&&+g_{12}[A^\dagger_{1}A_{2}e^{\mathrm{i}\Delta'_{12}t} +\!A^\dagger_{1}B_{2}e^{\mathrm{i}(\Delta'_{12} +\!\alpha_2)t}\notag\\
	&&+B^\dagger_{1}A_{2}e^{\mathrm{i}(\Delta'_{12} -\!\alpha_1)t}]\!\times\!e^{-\mathrm{i}\beta_1\cos(\nu_1t\!+\!\varphi_1(t))}\notag\\
	&&\times e^{\mathrm{i}\beta_2\cos(\nu_2t\!+\!\varphi_2(t))}\!+\!\mathrm{H.c.}\},
\end{eqnarray}
where $\Delta'_{ij}\!=\!\Delta_{ij}\!-\!\Delta'_{i}\!+\!\Delta'_{j}$ with $\Delta_{ij}\!=\!\omega_{qi}\!-\!\omega_{qj}$ being the  transition frequency difference between $\mathrm{T}_i$ and $\mathrm{T}_j$.   Eq. (\ref{HT1t}) describes two independent subspaces, i.e., the single-  and double-excitation subspaces, { as illustrated in   Fig. \ref{Figure.6}(c) and Fig. \ref{Figure.6}(d), respectively. These two subspaces can be selectively addressed by the frequencies of the driving fields.} By setting drive parameters $\nu_1\!=\!\Delta'_{1a}$, $\nu_2\!=\!\Delta'_{a2}$, $\varphi'_1(t)\!=\!\varphi_1(t)\!+\!\frac{\pi}{2}$, and $\varphi'_2(t)\!=\!\varphi_2(t)\!-\!\frac{\pi}{2}$, we can select the single-excitation subspace $\{|100\rangle_{1a2},|001\rangle_{1a2},|010\rangle_{1a2}\}$, which is $\Delta$-type coupled three-level system. To explicitly show this, we apply Jacobi-Anger expansion and ignore high-frequency oscillation terms, then we obtain an effective Hamiltonian
\begin{eqnarray}
	\label{Heff1}
	\mathcal{H}^{eff}_1(t)&&=\Delta'_1|0\rangle_\mathrm{L}\langle 0|\!+\!\Delta'_a|e\rangle_\mathrm{L}\langle e|\!+\!\Delta'_2|1\rangle_\mathrm{L}\langle 1|\notag\\
    &&+g_{1a}J_1(\beta_1)|0\rangle_\mathrm{L}\langle e|e^{-\mathrm{i}\varphi'_1(t)}\!+\!g_{a2}J_1(\beta_2)|1\rangle_\mathrm{L}\langle e|e^{\mathrm{i}\varphi'_2(t)}\notag\\
    &&+g_{12}J_1(\beta_1)J_1(\beta_2)|0\rangle_\mathrm{L}\langle 1|e^{-\mathrm{i}(\varphi'_1(t)\!+\!\varphi'_2(t))}\!+\!\mathrm{H.c.},
\end{eqnarray}
where $J_1(\beta_j)$ is Bessel functions of the first kind.

As Eqs. (\ref{Heff1}) and  (\ref{Htrans}) have the same mathematical form, the non-resonant coupled $\Delta$-type three-level system in Eq. (\ref{Htrans}) can be obtained, from which arbitrary single-qubit gate in Eq. (\ref{U01}) can be implemented, by setting parameters $\Delta'_1\!=\!-\frac{1}{2}\Delta_2\sin^2(\theta/2)$, $\Delta'_a\!=\!\frac{1}{2}\Delta_2$,  $\Delta'_2\!=\!-\frac{1}{2}\Delta_2 \cos^2(\theta/2)$,
$g_{1a}J_1(\beta_1)\!=\!\frac{1}{2}\Omega \sin(\theta/2)$,
$g_{a2}J_1(\beta_2)\!=\!\frac{1}{2}\Omega \cos(\theta/2)$,
$g_{12}J_1(\beta_1)J_1(\beta_2)\!=\!-\frac{1}{2}\Delta_2 \sin(\theta/2)\cos(\theta/2)$,
$\varphi'_1(t)\!=\!\phi_0(t)$, $\varphi'_2(t)\!=\!-\phi_1(t)$. For example,  to implement $R_x(\pi/2)$ in Eq. (\ref{U01}), we just need to set parameters $\Delta'_{j}$, $\beta_j$, $\nu_j$ and $\varphi_j$ according to the  parameters $\Delta_2$, $\Omega$, $\theta$ and $\phi_2$. Similarly,  other gates can also be implemented.

\subsection{Two-qubit gates case}

{As it is well know, the realization of universal quantum computation requires not only arbitrary single-qubit gates  but also a nontrivial two-qubit gate. And, the  two-logical-qubit DFS is $S_2\!=\!\{|100100\rangle_{1a23b4}\!=\!|00\rangle_\mathrm{L}, |100001\rangle_{1a23b4}\!=\!|01\rangle_\mathrm{L}, |001100\rangle_{1a23b4}\!=\!|10\rangle_\mathrm{L}, |001001\rangle_{1a23b4}\!=\!|11\rangle_\mathrm{L}\}$. Here, we choose to implement a controlled phase gate, which can also be accelerated by detuning adjustment. Remarkably, the controlled phase gate can be realized by coupling two adjacent transmon qubits from the two logical units, for example $\mathrm{T}_2$ and $\mathrm{T}_3$, as shown in Fig. \ref{Figure.6}(a).
Applying a driving filed on transmon  $\mathrm{T}_3$, i.e., $\omega_{q3}(t)\!=\!\omega_{q3}\!+\!\varepsilon_{3}\sin(\nu_{3}t\!+\!\varphi_{3}(t))$,  parametrically tunable coupling  can be achieved between transmons $\mathrm{T}_2$ and $\mathrm{T}_3$, and the coupling Hamiltonian is
\begin{eqnarray}
	\label{H2t}
	\mathcal{H}_2(t)&&=\sum_{j=2,3}{\left[\omega_{qj}|1\rangle_j\langle1|+(2\omega_{j}-\alpha_j)|2\rangle_j\langle 2|\right]}\notag\\
	&&+g_{23}(\sigma_{2}+\sigma^\dagger_{2})\otimes(\sigma_{3}+\sigma^\dagger_{3}).
\end{eqnarray}
In the interaction picture, under the rotating wave approximation, the interacting Hamiltonian of $\mathcal{H}_2(t)$ reads
\begin{eqnarray}
\label{HT2t1}
\mathcal{H}^T_2(t)&&=g_{23}[A^\dagger_{2}A_{3}e^{\mathrm{i}\Delta_{23}t} \!+\!A^\dagger_{2}B_{3}e^{\mathrm{i}(\Delta_{23}+\alpha_3)t}\notag\\
&&\!+\!B^\dagger_{2}A_{3}e^{\mathrm{i}(\Delta_{23}-\alpha_2)t}]\!\times\! e^{\mathrm{i}\beta_3\cos(\nu_3t+\varphi_3(t))}\!+\!\mathrm{H.c.}.
\end{eqnarray}
When the frequency of the driving field is set to $\nu_3\!=\!\Delta_{23}\!-\!\alpha_2\!-\!\Delta'_3$, with $\Delta'_3 \ll \{\Delta_{23}, \alpha_2\}$ being the adjustable detuning. By this setting, except for the $B^\dagger_{2}A_{3}$ term,  the rest terms in Eq. (\ref{HT2t1}) are all  of the high-frequency oscillating nature. Explicitly,  after applying Jacobi-Anger expansion and ignoring high-order terms,  $\mathcal{H}^T_2(t)$  reduces to
\begin{eqnarray}
	\label{HT2t2}
	\mathcal{H}^T_2(t)=ge^{-\mathrm{i}\varphi'(t)}|20\rangle_{23}\langle 11|e^{\mathrm{i}\Delta'_3 t}+\mathrm{H.c.},
\end{eqnarray}
where the effective coupling strength is $g\!=\!\sqrt{2}J_1(\beta_3)g_{23}$ and $\varphi'(t)\!=\!\varphi_3(t)\!-\!\pi/2$. After applying  a unitary transformation $U_2\!=\exp\{-\frac{\mathrm{i}}{2}\Delta'_3 t(|10\rangle_\mathrm{L}\langle 10|\!-\!|f\rangle_\mathrm{L}\langle f|)\}$, in the subspace $S_2'=\{|10\rangle_\mathrm{L},|f\rangle_\mathrm{L}\}$ with  $|f\rangle_\mathrm{L}\!=\!|002000\rangle_{1a23b4}$ being an auxiliary state, the reduced Hamiltonian  of $\mathcal{H}^T_2(t)$ becomes
\begin{eqnarray}
	\label{Heff22}
	\mathcal{H}^{\text{eff}}_2(t)=\begin{pmatrix} -\frac{1}{2}\Delta'_3& ge^{\mathrm{i}\varphi'(t)}\\ge^{-\mathrm{i}\varphi'(t)} &\frac{1}{2}\Delta'_3
\end{pmatrix}.
\end{eqnarray}
The implemented evolution operator from the above Hamiltonian is the same as the single-qubit gate case in Eq. (\ref{Ube}). In this case, the constraint on $\varphi'(t)$ under TOC will be reduced to $\varphi'\!(t)\!=\!\eta' t$, where $\eta'$ is a constant.
Explicitly, by setting $J' =\sqrt{g^2\!+\![(\Delta'_3\!-\!\eta')/2]^2}$ and $\chi'\!=\!2\tan^{-1}[2g/(\Delta'_3\!-\!\eta')]$, at the final time $\tau'$, the resulting evolution operator in the subspace $S_2'$ can be obtained as
\begin{eqnarray}
\label{Utau'1}
U(\tau')&&=e^{\mathrm{i}\frac{1}{2}\eta'\tau'\sigma'_z}\times\notag\\
&&\!\left(
\begin{array}{cccc}
\!\!\cos\xi'\!+\!\mathrm{i}\sin\xi' \cos\frac{\chi'}{2}            & \!-\mathrm{i}\sin\xi' \sin\frac{\chi'}{2} \\
\!-\mathrm{i}\sin\xi' \sin\frac{\chi'}{2} & \!\cos\xi'\!-\!\mathrm{i}\sin\xi' \cos\frac{\chi'}{2}\!\!
\end{array}
\right),
\end{eqnarray}
where $\xi'\!=\!J'\tau'$ and $\sigma'_z$ is Pauli operator for the subspace $S_2'$. Furthermore, to ensure that the cyclic evolution condition is satisfied, we here set $\xi'\!=\!\pi$, i.e., $\tau'\!=\!\pi/\sqrt{g^2\!+\![(\Delta'_3\!-\!\eta')/2]^2}$. In this way, our final evolution operator will be
\begin{eqnarray}
\label{Utau'2}
U(\tau')=e^{\mathrm{i}\gamma'}|10\rangle_\mathrm{L}\langle 10|+e^{-\mathrm{i}\gamma'}|f\rangle_\mathrm{L}\langle f|,
\end{eqnarray}	
where the total phase $\gamma\!=\!\pi\!+\!\frac{1}{2}\eta'\tau'$. Specifically, in the two-logical-qubit subspace $S_2$, the implemented gate will be
\begin{eqnarray}
	\label{Utau'}
	U(\tau')=\mathrm{diag}(1, 1, e^{\mathrm{i}\gamma'}, 1),
\end{eqnarray}
which is a non-trivial two-qubit gate, and it can be further accelerated by adjusting  $\Delta'_3$, similar to the single-qubit gate case.  }

\section{Conclusion}
In conclusion, we propose a NHQC scheme based on $\Delta$-type three-level system, combining with the TOC technique. Remarkably, by using the detuning adjustment, our proposal can greatly shorten the gate-time  within the hardware limitations, and thus has higher gate fidelities and stronger robustness than previous ones. Moreover, we present a simplified realization of our protocol, with DFS encoding, on a scalable 2D superconducting quantum circuit. Therefore, our scheme provides a promising alternation towards future large-scale fault-tolerant quantum computation.

\section*{Acknowledgements}

This work was supported by the Key-Area Research and Development Program of GuangDong Province (No. 2018B030326001), the National Natural Science Foundation of China (No. 11874156),
and the Science and Technology Program of Guangzhou (No. 2019050001).


\end{document}